\documentclass[twocolumn,prl,showpacs]{revtex4}

\usepackage{graphicx,amssymb,amsmath}
\usepackage[applemac]{inputenc}

\begin{document}

\title{Wave turbulence served up on a plate}
\author{Pablo Cobelli, Philippe Petitjeans}
\affiliation{Physique et M\'ecanique des Milieux H\'et\'erog\`enes, ESPCI \& CNRS, 10 rue Vauquelin, 75005 Paris, France.}
\author{Agn\`es Maurel}
\affiliation{Laboratoire Ondes et Acoustique, ESPCI \& CNRS, 10 rue Vauquelin, 75005 Paris, France.}
\author{Vincent Pagneux}
\affiliation{Laboratoire d'Acoustique de l'Universit\'e du Maine \& CNRS, Avenue Olivier Messiaen, 72085 Le Mans, France.}
\author{Nicolas Mordant}
\email[]{nmordant@ens.fr}
\affiliation{Laboratoire de Physique Statistique, Ecole Normale Sup\'erieure \& CNRS, 24 rue Lhomond, 75005 Paris, France.}


\begin{abstract}
Wave turbulence in a thin elastic plate is experimentally investigated. By using a Fourier transform profilometry technique, the deformation field of the plate surface is measured simultaneously in time and space. This enables us to compute the wavevector-frequency Fourier ($\mathbf k, \omega$) spectrum of the full space-time deformation velocity.  In the 3D ($\mathbf k, \omega$) space, we show that the energy of the motion is concentrated on a 2D surface that represents a nonlinear dispersion relation. This nonlinear dispersion relation is close to the linear dispersion relation. This validates the usual wavenumber-frequency change of variables used in many experimental studies of wave turbulence. The deviation from the linear dispersion, which increases with the input power of the forcing, is attributed to weak non linear effects. Our technique opens the way for many new extensive quantitative comparisons between theory and experiments of wave turbulence.
\end{abstract}

\maketitle

Wave turbulence is a state of waves in non-linear interaction as observed for a large variety of systems including Alfven waves in solar winds~\cite{Ng, Galtier}, ocean waves~\cite{Hasselmann}, non linear optics~\cite{Dyachenko} and superfluids~\cite{Lvov}. Similarly to the phenomenological theory of hydrodynamic turbulence, the so-called weak turbulence (WT) theory for wave turbulence predicts a Kolmogorov-Zakharov energy cascade~\cite{Zakharov}.  This analytical weak turbulence theory assumes the persistence of the space-time structure of the linear waves through the dispersion relation. Very few experimental studies have taken place and results show only partial agreement with theory~\cite{Denissenko,Falcon,Mordant,Boudaoud}. Furthermore, almost none of these experiments look beyond the analysis of measurement at a single point.
Here we report the analysis of the turbulence of bending waves on a shaken, thin elastic plate, a phenomenon used in theatres to simulate the sound of thunder. We are able to measure the fully resolved space-time dynamics of the deformation of the plate and we show that the energy is localized on a line in the wavenumber-frequency plane of the Fourier spectrum. This confirms the persistence of the space-time structure of waves which is the premise of weak turbulence theory. In addition, our system displays the phenomenology described by the theory and yet some of its predictions are not quantitatively fulfilled: the non-linear shift to the dispersion relation and the power spectrum do not obey the predicted scaling laws.

The theory of WT relies on the assumption of weak non linearity of waves. The latter induces a scale separation in the time evolution of the wave amplitude compared to the wave period and it provides a natural closure of the hierarchy of cumulants derived from the wave equation~\cite{Zakharov,Newell}. In contrast, no such closure can be exhibited for hydrodynamic turbulence. In particular, the WT theory of wave turbulence leads to a kinetic equation for the evolution of the energy spectrum of the waves. Stationary solutions are exhibited which corresponds to the Rayleigh-Jeans spectrum for systems in equilibrium and the Kolmogorov-Zakharov energy cascade for non-equilibrium systems. This prediction of the power spectrum density of the wave amplitude has been derived in many cases such as non linear optics, superfluids, gravity-capillary water waves, sound waves, Alfven waves, plasmas, oceanography, semiconductor lasers and bending waves in elastic plates~\cite{Zakharov,Newell,During}. There are a paucity of experiments specifically designed for wave turbulence and of those, most concern surface waves on liquids~\cite{Denissenko,Falcon}.
Our wave system consists of a thin steel plate on which elastic bending waves are excited by an electromagnetic vibrator. The dynamics of the plate follow the Föppl-Von Karman equations for the deformation:
 \begin{eqnarray}
&&\rho\frac{\partial^2\zeta}{\partial t^2}=-\frac{Eh^2}{12(1-\sigma^2)}\Delta^2\zeta+\{\zeta,\chi\}
\label{fvk1}\\
&&\frac{1}{E}\Delta^2\chi=-\frac{1}{2}\{\zeta,\zeta\}
\end{eqnarray}
where $\rho$ is the density, $\zeta$ the plate deformation, $h$ the plate thickness, $E$ the Young’s modulus, $\sigma$  the Poisson ratio, $\Delta$ the Laplacian operator, $\chi$ the stress function and $\{.,.\}$ is a bilinear differential operator~\cite{During}. Linearizing the first equation in~(\ref{fvk1}) provides the linear dispersion relation 
\begin{equation}
\omega_k=\sqrt{\frac{Eh^2}{12\rho(1-\sigma^2)}}k^2\, .
\label{rd}
\end{equation} 
The non linear term in (\ref{fvk1}) is due to the stretching of the plate and it is cubic in the wave amplitude. The WT theory has recently been applied to this case~\cite{During} and predicts a space Fourier spectrum of the amplitude of the waves 
\begin{equation}
E_\zeta(k)=C\frac{P^{1/3}}{(12(1-\sigma^2))^{1/6}}\frac{\ln^{1/3}(k^\star/k)}{\sqrt{E/\rho}\,\,k^3}\,
\end{equation} 
where $P$ is the average power input in the system from the applied forcing, $C$ is a number and $k^\star$ is a cut-off frequency. The one-point spectrum of the waves has been investigated experimentally~\cite{Mordant,Boudaoud} and has been shown not to obey the WT prediction in particular in its scaling in $P$ :
\begin{equation}
E(k)\propto \frac{P^{0.7}}{k^{4.2}}
\label{exp}
\end{equation}
Nevertheless, it displays a turbulent-like behaviour , i.e. a broadband spectrum, and the question is raised whether or not the disagreement with the theory is due to strongly non linear structures, to boundary condition effects or to some dissipative mechanism. 

\begin{figure}[htb]
\begin{center}
(a)\includegraphics[width=8cm]{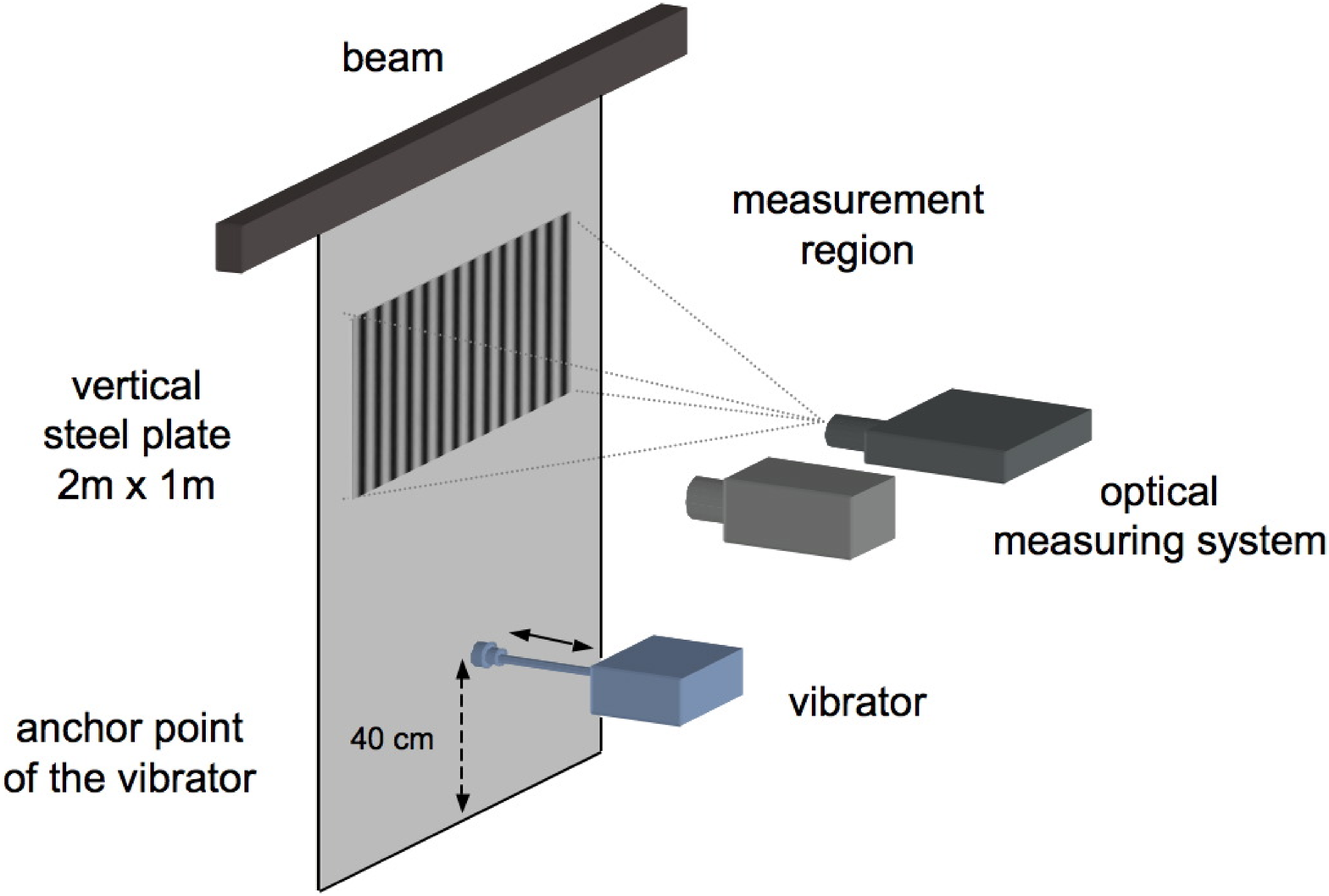}
(b)\includegraphics[width=8cm]{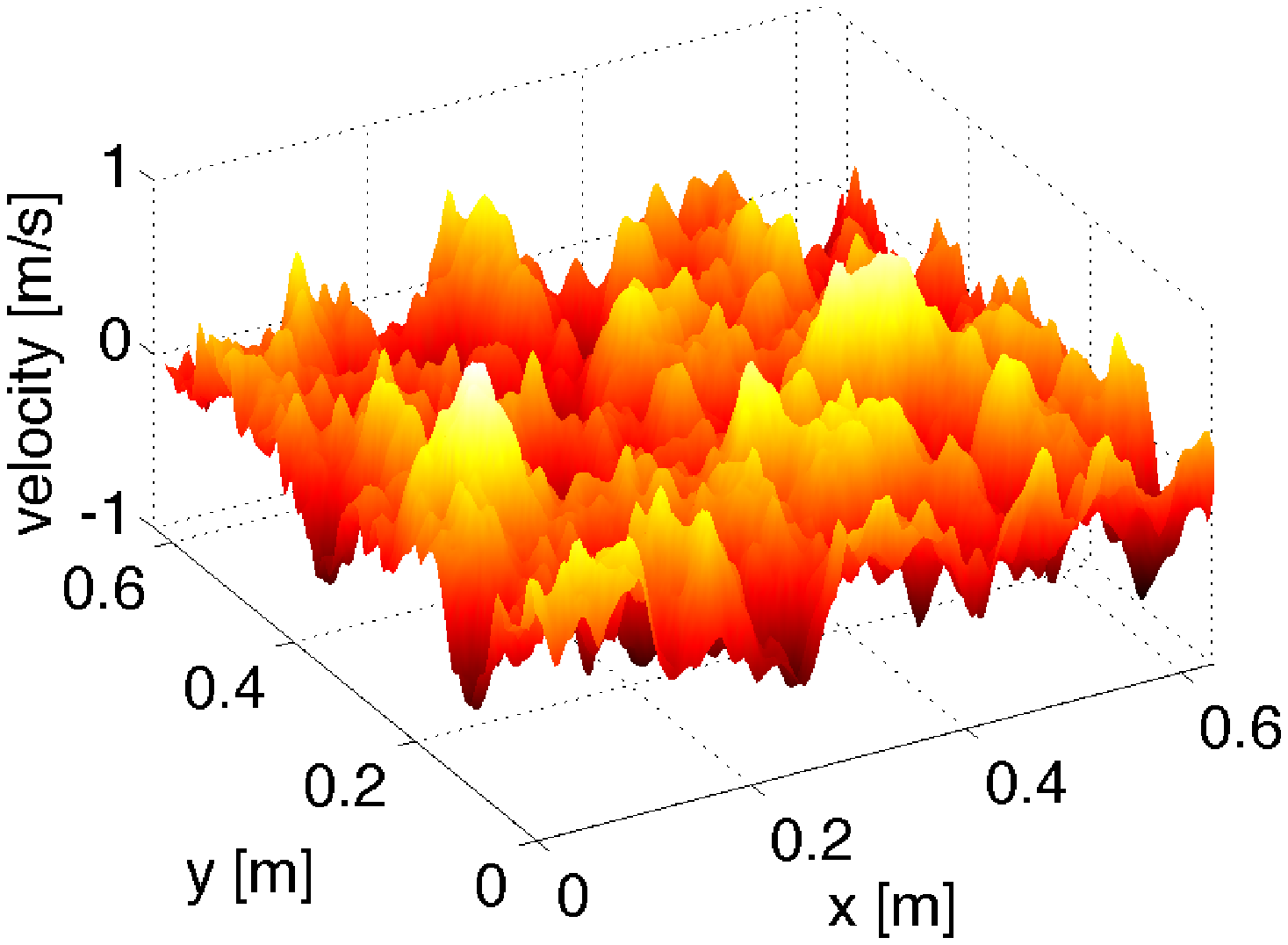}
\caption{(a) Sketch of the wave turbulence experiment. The specifically designed 2D mechanical system is made of a $2\times1$~m$^2$, 0.4~mm thick stainless steel plate held vertically and set in motion by an electromagnetic vibrator at 30~Hz. The  Fourier Transform profilometry is based on the projection of a sine intensity pattern by a high definition video projector. The deformed image is then recorded by a high speed camera (1300 or 2600~fps). (b) Example of measurement of the deformation velocity on a  63~cm by 62~cm area. }
\label{fig1}
\end{center}
\end{figure}
A sketch of the experimental setup is shown in fig.~\ref{fig1}(a). The plate is made of stainless steel and its size is 2 m by 1 m and 0.4 mm thick. Its is bolted on a I beam by one short end and is hanging under its own weight. An electromagnetic vibrator is anchored 40~cm from the bottom of the plate and excites the waves at 30~Hz with a varying amplitude. The vibrator is fitted with a FGP sensors force probe and Brüel \& Kjaer accelerometer to measure the input power $P$. A Fourier transform profilometry technique~\cite{Cobelli,Maurel} gives access to the temporal evolution of the deformation of the plate measured over a significant portion of its area. The principle is the following: a sine intensity pattern $I(x,y)\propto \sin(2\pi y/p)$ is projected on the surface of the plate by a videoprojector. The pattern is then recorded by a Phantom v9 high speed camera. The deformation of the plate induces a phase shift of the pattern recorded by the camera. The deformation of the plate is recovered by a 2D phase demodulation of each image in the movie~\cite{Maurel,Cobelli}. Movies are recorded either with $1000^2$ (resp. $800^2$) pixels at 1300 (resp. 2600) frames per seconds (fps). The configuration and the processing is similar to that of Cobelli et al.~\cite{Cobelli} with a distance of  $L=193$~cm from the projector to the plate  and a distance of $D=35$~cm between the optical axes. The normal velocity of the plate is obtained by differentiating the deformation movie in time. The field of view is about $71^2$~cm$^2$ at 1300~fps and $62^2$~cm$^2$ at 2600~fps. The spectra are calculated by performing a multidimensional Fourier transform without applying any windowing to preserve the localization of the energy in the Fourier space.
An example of the normal velocity of the plate is displayed in fig.~\ref{fig1}. 

\begin{figure}[htb]
\begin{center}
(a)\includegraphics[width=8cm]{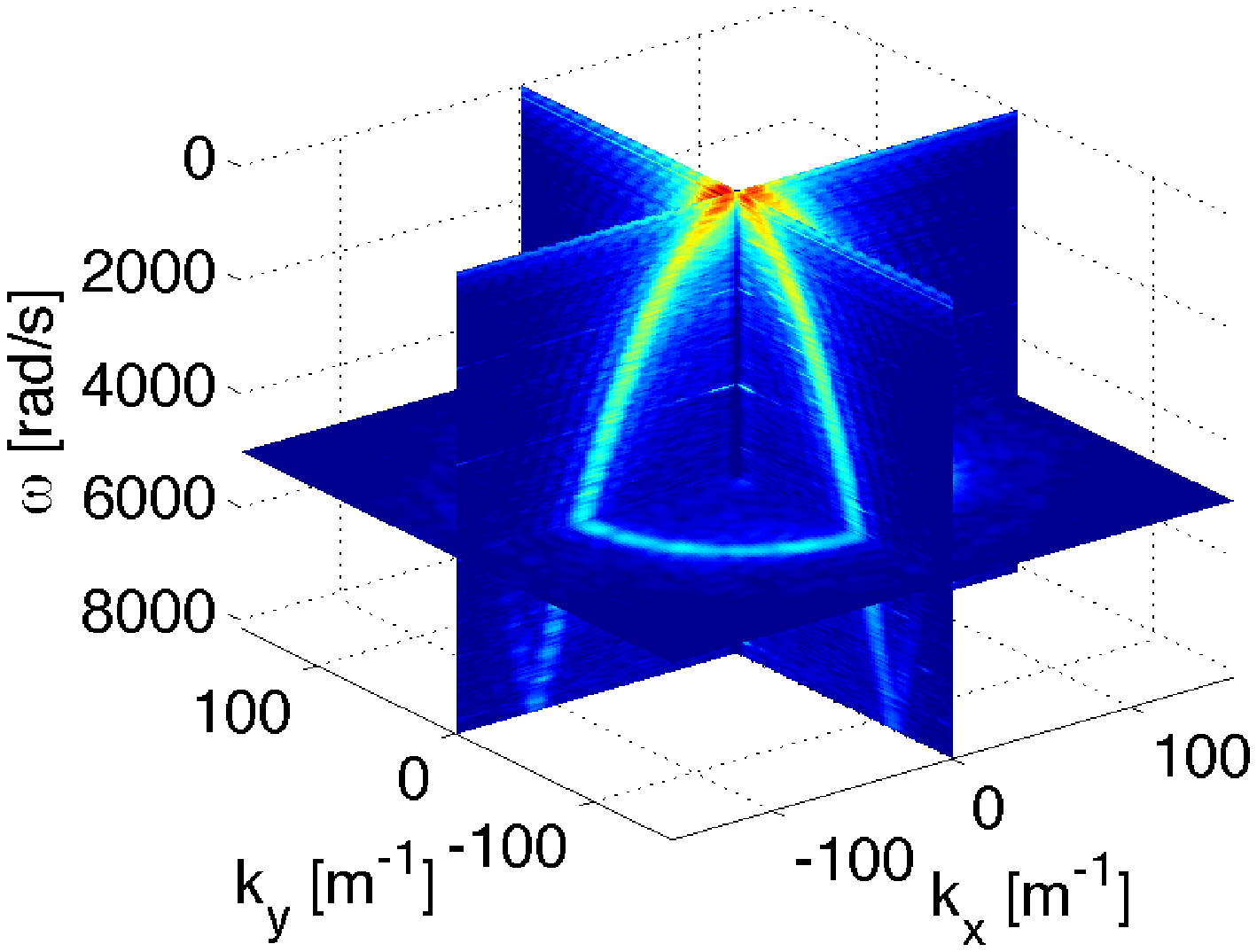}
(b)\includegraphics[width=8cm]{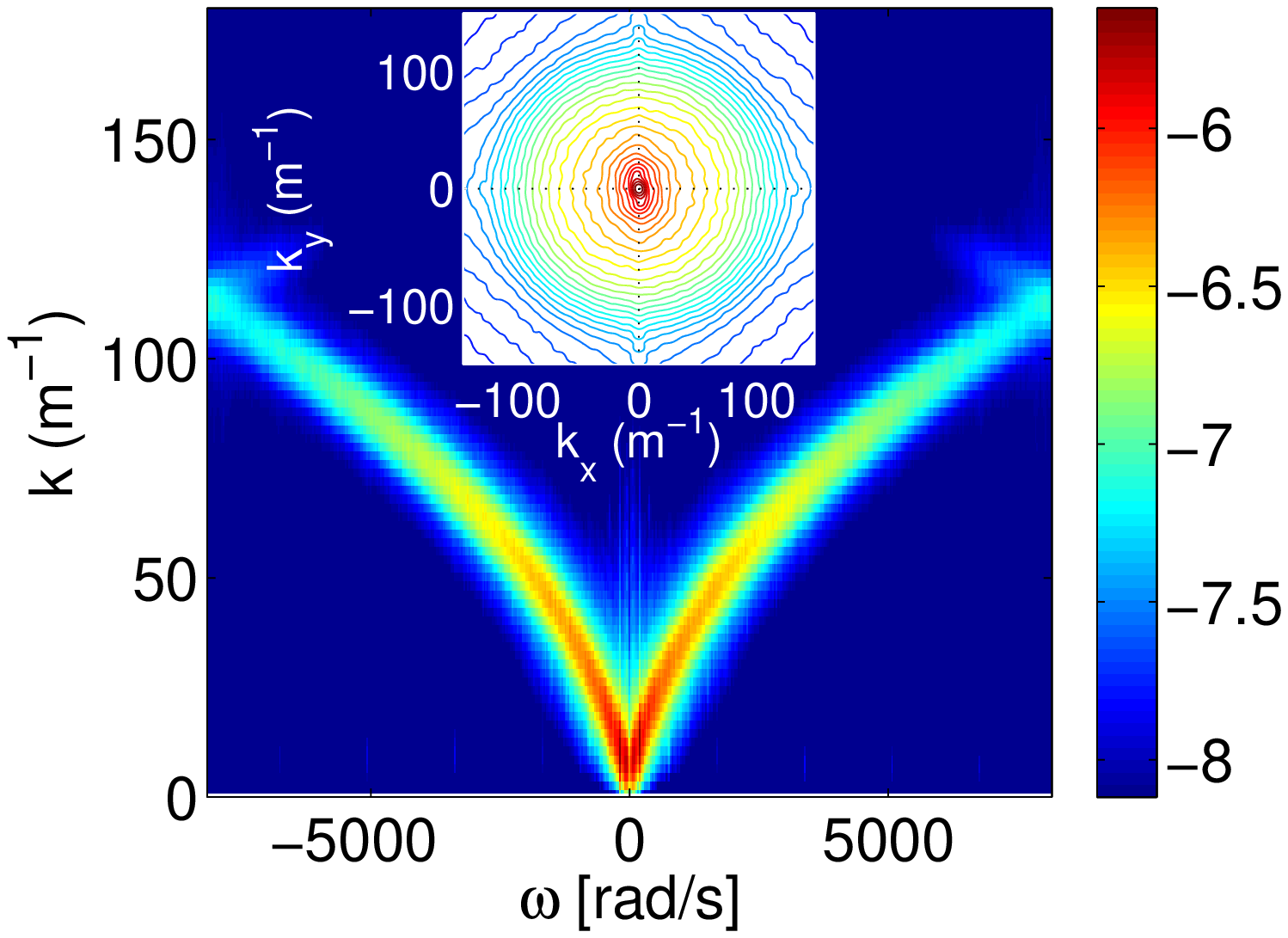}
\caption{(a) Space-time spectrum $E(\mathbf k,\omega)$ of the deformation velocity (colors are log scaled). The cuts are located at $k_x=0$, $k_y=0$ and $\omega=5000$~rad/s. The energy is localized on a surface in the $(\mathbf k,\omega)$ space which confirms that the turbulent motion is due to a non linear superposition of waves following a dispersion relation. (b) Spectrum $E(k,\omega)$ obtained from $E(\mathbf k,\omega)$ by integrating over the direction of $\mathbf k$. Insert: space spectrum $E(\mathbf k)$ computed from $E(\mathbf k,\omega)$ by summing over the frequencies. Contours are log scaled in both plots.}
\label{fig2}
\end{center}
\end{figure}
The full space-time Fourier spectrum (shown in fig. \ref{fig2}(a)) of the deformation $E(\mathbf k,\omega)$ (a function of both the wave vector $\mathbf k$ and the frequency $\omega$) is constructed from the movie of the deformation velocity. The striking feature is the localization of the energy in the vicinity of a surface showing that the motion is a non linear superposition of waves following a dispersion relation $\omega=f(\mathbf k)$, close to the linear dispersion relation. This is the first experimental observation of such a space-time spectrum in wave turbulence. 
In addition to the full space-time spectrum, we can analyse the space spectrum $E(\mathbf k)$, as displayed in fig.~\ref{fig2}(b). The isocontours for large wave numbers are circles, revealing the isotropy of the spectrum in this regime. The anisotropic response to the forcing is visible at low wave numbers. This behaviour is expected in the phenomenology of the Kolmogorov cascade of energy and is evidenced here: After a few steps in the cascade, the anisotropy of the forcing is forgotten down to the small scales at which the dissipation is dominant. 
Owing to the isotropy of the spectrum, in fig.~\ref{fig2}(b) we show the spectrum $E(k=\|\mathbf k\|,\omega)$ obtained by integrating $E(\mathbf k,\omega)$ over all the directions of the wave vector. The localisation of the energy appears as a line in the $(k,\omega)$ plane. The width of this line is close to the inverse of the image size, which indicates that the localization of the energy in our measurement is actually limited by the resolution of the Fourier transform due to the finite size of the plate. At low frequency, the injection of energy corresponds to a peak on the energy line: even though the forcing is localized in space, its monochromatic nature (at 188 rad/s) makes it local in the $(k,\omega)$ plane. The forcing operates effectively at low frequency and at low wavenumber as is expected in the phenomenology of the Kolmogorov-Zakharov cascade.

\begin{figure}[htb]
\begin{center}
\includegraphics[width=9cm]{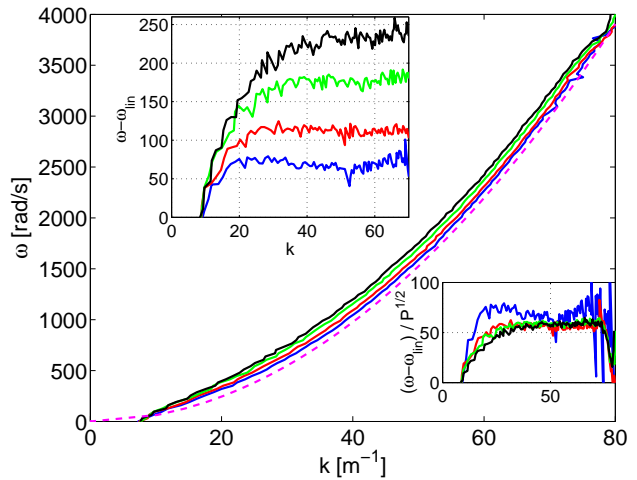}
\caption{Non linear dispersion relation $\omega(k)$ computed from the line of maximum energy in the space-time spectrum of the plate deformation velocity for various input power $P$ of the forcing (from bottom to top $P^{1/2}=1$, 2, 3, 4 in arbitrary units). The dashed line is the linear dispersion relation (\ref{rd}). A systematic shift is observed which increases with the forcing power. The top insert shows the deviation from the linear dispersion. At high wave numbers, the shift is seen to be independent of $k$. The bottom insert shows the shift normalized by $P^{1/2}$. }
\label{fig3}
\end{center}
\end{figure}
The concentrated line of energy in the spectrum $E(k,\omega)$ allows the dispersion relation to be computed; it is extracted by computing the position of the crest of the energy line at each frequency and is displayed in fig.~\ref{fig3} for various values of the forcing $P$. The dispersion relation remains close to the linear dispersion relation with a small but systematic shift. This provides strong evidence that our system is indeed weakly non linear. Thus, the quantitative disagreement between  the one point spectrum and the WT theory prediction~\cite{Mordant,Boudaoud} cannot be attributed to the existence of  strongly non linear structures.  Instead, the disagreement is proposed to be attributed to the dissipative mechanisms which are believed to exist at all scales rather than being present at only small scales; hence the Kolmogorov-Zakharov cascade is “leaking”~\cite{Josserand}.

Figure~\ref{fig3} allows us to quantify the departure of the observed dispersion relation from the linear one. Notably, it is shown that the correction increases with the power input $P$ with a behaviour close to $P^{1/2}$ behaviour. It is also observed to be constant at high wavenumbers so that the various dispersion relations are parallel to one another. This behaviour is of particular interest when compared with the WT prediction. Indeed, it is expected that the departure of the dispersion relation from the linear one has a power law scaling in $P$ which is identical to the scaling for the energy spectrum~\cite{Rica}. Our experimental measurements confirm this prediction: the exponent close to $1/2$ in P is common to both departure from linearity in the dispersion relation and also to the energy spectrum.

\begin{figure}[htb]
\begin{center}
\includegraphics[width=8cm]{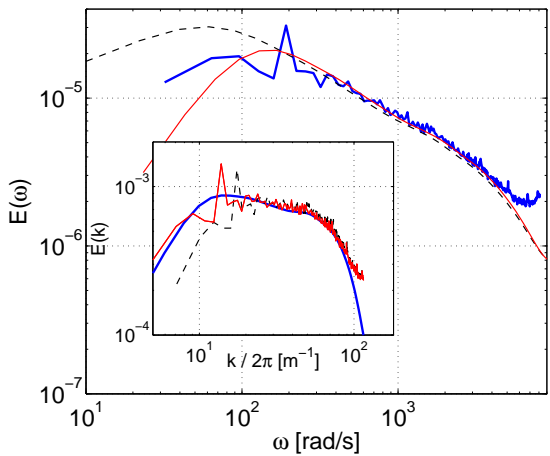}
\caption{Test of the validity of the change of variable $k\leftrightarrow \omega$ using the dispersion relation. Solid blue line: direct measurement of the time spectrum $E(\omega)$ -- solid red (resp black dashed) line: $E(\omega)$ computed from the space spectrum $E(k)$ using the nonlinear (resp. linear) dispersion relation to change variable. Insert: same for the space spectrum $E(k)$. Solid blue line: direct estimation of $E(k)$ -- red (resp black dashed) line: $E(k)$ computed from $E(\omega)$  by the change of variable using the nonlinear (resp. linear) dispersion relation.}
\label{fig4}
\end{center}
\end{figure}
The application of the WT turbulence theory is often restricted to the prediction of the Kolmogorov-Zakharov space spectrum $E(k)$. Although some experiments directly measure the space spectrum~\cite{Wright,Snouck}, it is often easier to measure the motion at one given point as a function of time. In that case, only the time spectrum $E(\omega)$ can be estimated. To compare with the theory, the space spectrum is determined by using the dispersion relation to obtain $E(k)$ from $E(\omega)$~\cite{Denissenko,Falcon,Henry}. This approach was used to deduce the law in Eq. (\ref{exp})~\cite{Mordant,Boudaoud}. We can independently estimate both $E(\omega)$ and $E(k)$ directly. We can assess the validity of the change of variables technique above (via the linear or non linear dispersion relation). The comparison of the various cases is shown in fig.~\ref{fig4}. Both dispersion relations allow us to reproduce the inertial range, with a better agreement shown when using the non linear relation. The large time or length scales are well reproduced only when using the non linear dispersion relation. This validates the usual change of variables when the non linearity is weak. 

Our experimental approach of wave turbulence reveals the main features of the weakly coupled waves that can be usefully compared with weak turbulence theory. Until now, only the spectrum $E(k)$ has been compared with theoretical prediction. However, WT theory can go far beyond spectra predictions: it gives quantitative predictions for multipoint statistics. Our present study confirms and quantifies the weakly nonlinear behaviour of the waves comprising the turbulent cascade. It confirms that the scaling law in the supplied power P is the same for the departure from the linear dispersion relation and also for the energy spectrum. Overcoming the discrepancy between experiments and theory claimed previously, we have shown some agreement between experimental results and WT theory. We anticipate that this experiment will allow precise and quantitative comparisons with theoretical investigations of wave turbulence~\cite{Zakharov,Kartashova,Philips} of prime importance for the large number of turbulent systems in which extensive measurements are out of reach.

\begin{acknowledgments}
N.M. thanks Agence Nationale de la Recherche for its funding under grant TURBONDE BLAN07-3-197846.
\end{acknowledgments}

\bibliography{cobelli}

 \end{document}